\documentclass{ws-ijmpb}
\usepackage{graphicx}

\newcommand{\fref}[1]{Fig.~\ref{#1}}
\newcommand{\phd}{{\protect \vphantom{\dagger}}}
\newcommand{\kf}{k_F}

\def \CDW{{\rm CDW}}

\def \SDW{{\rm SDW}}
\def \SS{{\rm SS}}

\def \TS{{\rm TS}}

\newcommand{\hc}{{\rm h.c.} }
\newcommand{\down}{\downarrow}
\newcommand{\up}{\uparrow}

\newcommand{\normfig}{0.7\textwidth}
\newcommand{\largefig}{0.8\textwidth}

\begin{document}

\markboth{M. A. Cazalilla, A. F. Ho and T. Giamarchi} {Deconfinement
and cold atoms in optical lattices}

%
\catchline{}{}{}{}{}
%

\title{Deconfinement and cold atoms in optical lattices}

\author{M. A. Cazalilla}

\address{Donostia Int'l Physics Center (DIPC), Manuel de Lardizabal,
4. 20018-Donostia, Spain.\\
waxcagum@sq.ehu.es}

\author{A. F. Ho}
\address{School of Physics and Astronomy,  The University of Birmingham,
Edgbaston, Birmingham B15 2TT, UK.\footnote{Present address: CMTH
Group, Dept. Physics, Blackett Laboratory,
Imperial College, Prince Consort Road, London SW7 2BW, UK.}\\
andrew.f.ho@imperial.ac.uk}

\author{T. Giamarchi}
\address{University of Geneva, 24 Quai Enerst-Ansermet, CH-1211
Geneva 4, Switzerland.\\
Thierry.Giamarchi@physics.unige.ch}

\maketitle

\begin{history}
\received{Day Month Year}
\revised{Day Month Year}
\end{history}

\begin{abstract}
Despite the fact that by now one dimensional and three dimensional
systems of interacting particles are reasonably well understood,
very little is known on how to go from the one dimensional physics
to the three dimensional one. This is in particular true in a
quasi-one dimensional geometry where the hopping of particles
between one dimensional chains or tubes can lead to a dimensional
crossover between a Luttinger liquid and more conventional high
dimensional states. Such a situation is relevant to many physical
systems. Recently cold atoms in optical traps have provided a unique
and controllable system in which to investigate this physics. We
thus analyze a system made of coupled one dimensional tubes of
interacting fermions. We explore the observable consequences, such
as the phase diagram for isolated tubes, and the possibility to
realize unusual superfluid phases in coupled tubes systems.
\end{abstract}


\section{Introduction}

As is well known, interactions have drastic effects on the behavior
of bosonic and fermionic systems, and change drastically their
properties compared to those of noninteracting particles.
Understanding the properties of such strongly correlated systems is
a particularly challenging problem. The effects of interaction are
also considerably enhanced when the dimension of the system is
reduced. Very naturally this quest for strongly correlated systems
has thus led to investigations of low dimensional systems, in
particular in condensed matter systems.  Among these low dimensional
systems, one dimensional ones play a special role. In such systems,
interaction effects are particularly strong since there is no way
for particles to avoid each other. For fermions interactions are
known to change drastically the properties compared to the canonical
ones of a Fermi liquid. The one dimensional interacting fermionic
system is indeed one of the very few solvable case of a non-fermi
liquid, known as a Luttinger liquid\cite{giamarchi_book_1d}. For
boson systems, here again the interactions lead to properties quite
different from the ones of weakly interacting bosons in one
dimension, and that are much closer to the ones of fermions: the
interacting one dimensional bosonic systems is again a Luttinger
liquid.

For fermions the progress in material science and nanotechnology
have recently made possible to realize such one dimensional systems.
Another class of materials in which such low dimensional systems
could be achieved with an unprecedented level of control has been
recently provided by cold
atoms\cite{greiner_mott_bec,stoferle_tonks_optical}. Indeed cold
atoms not only provide the possibility to realize both bosonic and
fermionic low dimensional systems, but the interactions and kinetic
energy can be controlled at will using optical lattices and Feshbach
resonances\cite{inouye_feshbach_resonances_bosons,timmermans_feshbach_theory,holland_feshbach_theory}.

Such systems allow also to tackle another question that is of
importance for a large class of materials. Since the effects of
interactions vary enormously with the dimension, it is crucial to
understand how one goes from the one dimensional behavior to a more
conventional three dimensional one. Understanding such dimensional
crossovers is crucially important for materials such as the organic
superconductors\cite{giamarchi_review_chemrev}, and even more
complicated anisotropic materials such as the high Tc
superconductors. Cold atomic systems thus provide a unique system
for which these questions could be investigated. This can be done by
studying the properties of quasi-one dimensional systems made of
many one dimensional tubes coupled together. Such systems have been
both investigated\cite{ho_deconfinement_coldatoms} and
realized\cite{stoferle_tonks_optical} for the case of bosons, and it
is now possible to achieve similar trapping in optical lattices for
the fermionic case as well\cite{kohl_fermions_3d}. In these notes we
thus focuss on such fermionic systems. We show how the peculiarity
of cold atoms leads to interesting properties already at the level
of an isolated 1D system. We then examine in details the effect of
the coupling between the tubes and shows that such systems can be
potential realizations of triplet superconductors. We expand in
these notes on Ref.~\refcite{cazalilla_fermions_1d} and refer the
reader to this paper for technical details and more specialized
references.

\section{Fermions, one dimension}

\begin{figure}
\begin{center}
\includegraphics[width=\largefig]{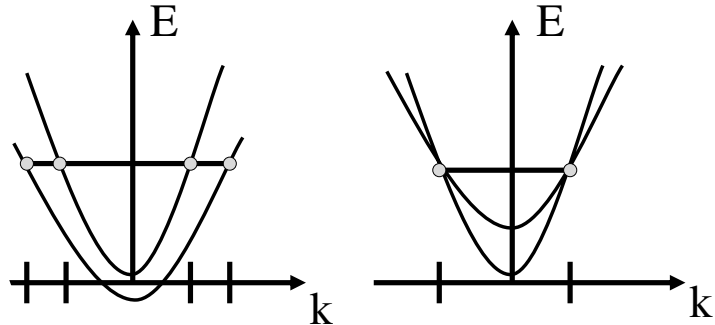}
\caption{Relation dispersion $E(k)$ for a system with spin dependent
hopping. Left: in the traditional case there are spin relaxation
phenomena ensuring the equilibrium of the two spin species. There is
thus a unique chemical potential for both species. This implies the
presence of four fermi points. Right: for cold atoms, there is no
such relaxation and the number of spin up and down are separately
conserved. For an equal number of spin up and down, there is thus
only two Fermi points since the two fermi momenta ${\kf}_\up =
{\kf}_\down$. \label{fig:potchem}}
\end{center}
\end{figure}

Let us first consider interacting one dimensional fermions. We use
the following extension of the traditional Hubbard model
\begin{equation} \label{eq:hamstart}
H = -\sum_{\sigma, m} t_{\sigma}   \left( c^{\dag}_{\sigma  m}
c_{\sigma  m+1} + \hc \right) +  U\sum_{m} n_{\uparrow  m}
n_{\downarrow  m}. \label{Ham}
\end{equation}
This Hamiltonian describes a 1D Fermi gas on a spin dependent
periodic potential. $t_\sigma$ is the hopping from site $m$ to site
$m+1$ for the spin species $\sigma$ and $U$ the local interaction.
In the cold atoms context the spin index refers to two hyperfine
states, or  two different types of atoms ({\it e.g.} $^6$Li and
$^{40}$K). Even though there may be no \emph{true} spin symmetry, we
use the spin  language to describe this binary mixture. It is
important to point out differences between (\ref{eq:hamstart}) and
the standard Hubbard model. First because the system is in an
optical lattice it is possible to control separately the hopping of
the up and down particles. As we discuss below this allows for a
richer phase diagram than the one of the standard Hubbard model.
Second, and most importantly, there is no mechanism in the cold atom
context that allows to relax the ``spin'' orientation, since this
corresponds to two different isospin states. As a result the number
of spin up and spin down is \emph{separately} conserved. This is at
strong variance with a standard condensed matter situation where one
generally impose a \emph{unique} chemical potential for the spin up
and spin down particles. The important difference, as shown in
\fref{fig:potchem}, is that there are, for the case of equal number
of spin up and spin down species, only two Fermi points instead of
four.
For such a system, the various operators corresponding to the
different types of order in the tubes are of two
types\cite{giamarchi_book_1d}. There are charge density or spin
density order described by
\begin{eqnarray}\label{eq:opdensspin}
 O_{\rho}(x) &=& \sum_{\sigma,\sigma'}
 \psi^\dagger_{\sigma}(x) \delta_{\sigma,\sigma'}
 \psi^\phd_{\sigma'}(x) \\
 O_{\sigma}^a(x) &=& \sum_{\sigma,\sigma'}
 \psi^\dagger_{\sigma}(x) \sigma^a_{\sigma,\sigma'}
 \psi^\phd_{\sigma'}(x)
\end{eqnarray}
where $\sigma^x$, $\sigma^y$, $\sigma^z$ are the Pauli matrices The
operator $O_{\rho}$ is simply the total density, whereas
$O_{\sigma}^a(x)$ measures the spin density (along the direction
$a=x,y,z$). We have used the notation in the continuum $c_m \to
\psi(x)$ as the operator destroying a fermion at point $x$. The
density contains a $q\sim 0$ component and a $q\sim 2\kf$ one. The
operators giving the $2\kf$ modulation of the charge or spin density
are
\begin{eqnarray}\label{eq:chspinop}
 O_{\CDW}(x) &=& \psi^\dagger_{R\up}\psi^\phd_{L\up}(x)+\psi^\dagger_{R\down}\psi^\phd_{L\down}(x)
 = \frac{e^{-2i\kf x}}{\pi\alpha}e^{i\sqrt2\phi_\rho}\cos(\sqrt2\phi_\sigma) \\
 O_{\SDW}^x(x) &=& \psi^\dagger_{R\up}\psi^\phd_{L\down}(x)+\psi^\dagger_{R\down}\psi^\phd_{L\up}(x)
 = \frac{e^{-2i\kf x}}{\pi\alpha}e^{i\sqrt2\phi_\rho}\cos(\sqrt2\theta_\sigma) \\
 O_{\SDW}^y(x) &=&
 -i(\psi^\dagger_{R\up}\psi^\phd_{L\down}(x)-\psi^\dagger_{R\down}\psi^\phd_{L\up}(x))
 = \frac{-e^{-2i\kf x}}{\pi\alpha}e^{i\sqrt2\phi_\rho}\sin(\sqrt2\theta_\sigma) \\
 O_{\SDW}^z(x) &=& \psi^\dagger_{R\up}\psi^\phd_{L\up}(x)-\psi^\dagger_{R\down}\psi^\phd_{L\down}(x)
 = \frac{e^{-2i\kf x}}{\pi\alpha}e^{i\sqrt2\phi_\rho}i\sin(\sqrt2\phi_\sigma)
\end{eqnarray}
where we have used the notation $\psi^\dagger_R$ (resp.
$\psi^\dagger_L$) to denote the operator creating a fermion with a
momentum close to $+\kf$ (resp. $-\kf$). These operators have a
simple representation in terms of continuous field
$\phi_{\rho,\sigma}$ (resp. $\theta_{\rho,\sigma}$) describing the
collective excitations of charge and spin densities (resp. current
densities). This representation is the so-called bosonization
representation and is very useful to identify the types of order in
the system. For more details on this representation see e.g.
Ref.~\refcite{giamarchi_book_1d}. Here we will only use that $\phi$
and $\nabla\theta$ have canonical conjugation commutation relations,
so order in one of the field implies exponentially decreasing
correlations in the other. Besides the above mentioned types of
order, other instability in the system are the superfluid
instabilities of the singlet or triplet type
\begin{eqnarray}\label{eq:decaydenslut}
 O_{\SS}(x) &=& \sum_{\sigma,\sigma'} \sigma
 \psi^\dagger_{R,\sigma}(x) \delta_{\sigma,\sigma'}
 \psi^\dagger_{L,-\sigma'}(x)\\
 O_{\TS}^a(x) &=& \sum_{\sigma,\sigma'} \sigma
 \psi^\dagger_{R,\sigma}(x) \sigma^a_{\sigma,\sigma'}
 \psi^\dagger_{L,-\sigma'}(x)
\end{eqnarray}
where SS denotes singlet pairing whereas TS is triplet pairing.
These operators describe paring with zero total momentum. Other
pairings are of course possible but are usually less relevant. These
operators become
\begin{eqnarray}\label{eq:superconop}
 O_{\SS}(x) &=& \psi^\dagger_{R\up}\psi^\dagger_{L\down}(x)
 +\psi^\dagger_{L\up}\psi^\dagger_{R\down}(x)
 = \frac{1}{\pi\alpha}e^{-i\sqrt2\theta_\rho}\cos(\sqrt2\phi_\sigma)
 \\
 O_{\TS}^x(x) &=& \psi^\dagger_{R\up}\psi^\dagger_{L\up}(x)
 +\psi^\dagger_{L\down}\psi^\dagger_{R\down}(x)
 = \frac{1}{\pi\alpha}e^{-i\sqrt2\theta_\rho}\cos(\sqrt2\theta_\sigma)
 \\
 O_{\TS}^y(x) &=&
 -i(\psi^\dagger_{R\up}\psi^\dagger_{L\up}(x)-\psi^\dagger_{L\down}\psi^\dagger_{R\down}(x))
 = \frac{-1}{\pi\alpha}e^{-i\sqrt2\theta_\rho}\sin(\sqrt2\theta_\sigma)
 \\
 O_{\TS}^z(x) &=& \psi^\dagger_{R\up}\psi^\dagger_{L\down}(x)-\psi^\dagger_{L\up}\psi^\dagger_{R\down}(x)
 = \frac{e^{2i\kf x}}{\pi\alpha}e^{-i\sqrt2\theta_\rho}\sin(\sqrt2\phi_\sigma)
\end{eqnarray}

We will not dwell on the solution\cite{cazalilla_fermions_1d} of the
model here and will only quote the results. Compared to the standard
case of the Hubbard model the main effect of the spin dependent
hopping is to open a spin gap both for repulsive and attractive
interactions (the standard Hubbard model being massless for
repulsive interactions). For attractive interactions the spin gap
corresponds to the formation of singlets. In the bosonization
language this corresponds to the order $\phi_\sigma\to 0$.
Correlation functions containing $\theta_\sigma$ or
$\sin(\sqrt2\phi_\sigma)$ thus decay exponentially to zero. The
leading instabilities are thus the charge density wave (CDW) order
or a singlet superfluid (SS) instability. Physically it means that
fermions of opposite spin pair and that these pairs behave roughly
as bosons and can then condense to give a superfluid (SS) or
crystallize to give a charge density wave (CDW).

\begin{figure}
\begin{center}
\includegraphics[width=\normfig]{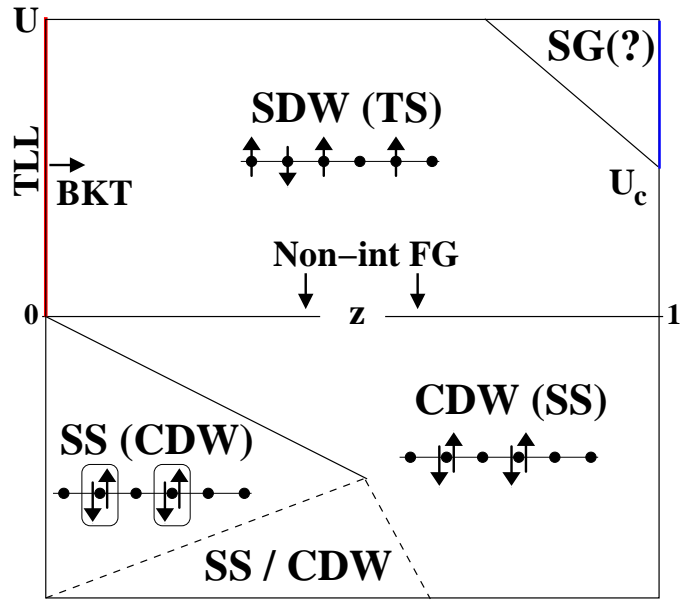}
\caption{Schematic phase diagram for the model in Eq. (\ref{Ham})
with equal number of spin up and down fermions away from
half-filling. The interaction strength is $U$ and $z = |t_{\uparrow}
- t_{\downarrow}|/ (t_{\uparrow}+t_{\downarrow})$. All phases (SDW:
spin density wave, CDW: charge density wave, SS singlet superfluid,
TS triplet superfluid)   exhibit a spin gap $\Delta_s$ (however,
$\Delta_s = 0$ for $U = 0$  and $z = 0$ with $U > 0$. A cartoon of
the type of order characterizing each phase is also shown. In the
area between dashed lines the dominant order (either CDW or SS)
depends on the lattice filling. In the SG phase, spin up and down
fermions are segregated (demixed). \label{fig1}}
\end{center}
\end{figure}

For repulsive interactions the hopping difference induce an order in
$\phi_\sigma\to \pi/\sqrt8$. Correlation functions containing
$\theta_\sigma$ or $\cos(\sqrt2\phi_\sigma)$ decay exponentially to
zero while $\sin(\sqrt2\phi_\sigma)$ tends to a constant. The
leading instabilities are thus a spin density wave order along the
$z$ direction ($\SDW_z$) or a triplet superfluid instability along
$z$ ($TS_z$). Which one of these instabilities dominates is dictated
by the decay of the correlations of the charge sector that remains
massless. For repulsive interactions, the $\SDW_z$ one dominates,
while the $\TS_z$ one decays with a larger power law and is thus
subdominant. A summary of the phase diagram is given in \fref{fig1}.
For repulsive interactions the spin sector is thus still massive but
the spin gap does not correspond to the formation of singlets. On
the contrary it is due to a breaking of the spin rotation symmetry
with formation of an ising-like antiferromagnetic order along the
$z$ direction. Note that even if the spin sector is massive, the
spin density wave order is not perfect (except at a commensurate
filling of exactly one fermion per site) due to the fluctuations in
the charge sector. Contrarily to the case of the Hubbard model where
the SDW order in the three directions decays in a similar way (due
to the spin rotation symmetry) here the SDW$_z$ order decays as a
power law because of charge fluctuations while SDW$_{x,y}$ order
decays exponentially fast because of the presence of the gap in the
spin sector. We refer the reader to
Ref.~\refcite{cazalilla_fermions_1d} for more details on the various
correlation functions and ways to probe the existence of such a gap
by Raman spectroscopy.

\section{Coupled Fermionic tubes}

Let us now turn to the case of coupled tubes. The coupling between
the tubes can be described by the single particle hopping term
\begin{equation} \label{eq:hperp}
 H_\perp = -t_\perp \sum_{\langle\alpha,\beta\rangle} \sum_{m,\sigma} c^\dagger_{\sigma,m,\alpha} c_{\sigma,m,\beta}
\end{equation}
where $\alpha$ and $\beta$ are the tube indexes and $\langle
\alpha,\beta \rangle$ denotes nearest neighbor tubes. Because for
cold atomic gases the interactions are short range, there is no need
to take into account interactions between different tubes. If the
single particle hopping (\ref{eq:hperp}) is of the same order of
magnitude than the ones in (\ref{eq:hamstart}) then the system is a
three dimensional system, while if $t_\perp = 0$ the tubes are
uncoupled. The Hamiltonian (\ref{eq:hperp}) is thus able to describe
the dimensional crossover between these two situations. For bosons
the effects of such a term have been investigated in
Ref.~\refcite{ho_deconfinement_coldatoms}. For fermions treating
such a single particle hopping is an extremely challenging problem.
Indeed, contrarily to the case of boson, the average of a single
fermion operator does not exist, and thus the term (\ref{eq:hperp})
cannot be decoupled simply. Various approximations have thus been
used to tackle this term and we refer the reader to
Ref.~\refcite{biermann_oned_crossover_review,giamarchi_review_chemrev,giamarchi_book_1d}
for details and further references.

However for the particular case of the fermionic system investigated
in these notes, one is in a much more favorable situation because of
the presence of the spin gap. In the case of attractive
interactions, the fermions form singlet pairs, that essentially
behave as bosons. One is thus essentially led back to the bosonic
case of Ref.~\refcite{ho_deconfinement_coldatoms}. We thus focus
here on the more interesting repulsive case. Because of the presence
of the spin gap the situation is now quite different to the one
where each tube is described by the simple Hubbard model. In the
latter case the spin excitations are massless and the single
particle correlations decay as power laws
\begin{equation}
 \langle T_\tau c_{\alpha,\sigma,}(x) c^\dagger_{\alpha,\sigma}(0)
 \rangle \propto \left(\frac1{x}\right)^{\frac14[K_\rho + 1/K_\rho] + \frac12}
\end{equation}
where $K_\rho$ is an interaction dependent parameter (the Luttinger
liquid parameter) characterizing the isolated tube. In that case a
simple scaling analysis to second order in perturbation in the
single particle hopping term (\ref{eq:hperp}) shows that it scales
as
\begin{equation}
 t_\perp^2 L^{4- \frac12[K_\rho + 1/K_\rho] - 1}
\end{equation}
and is thus in general a relevant perturbation, driving the system
away from the isolated tube fixed point. However in the presence of
a spin gap the \emph{single particle} excitations now decay
exponentially
\begin{equation}
 \langle T_\tau c_{\alpha,\sigma,}(x) c^\dagger_{\alpha,\sigma}(0)
 \rangle \propto e^{-|x|/\xi}
\end{equation}
where $\xi$ is the correlation length induced by the presence of the
spin gap (typically of order $\xi = v_F/\Delta_s$ if $v_F$ is the
Fermi velocity). The single particle hopping is now an irrelevant
perturbation to (\ref{eq:hamstart}). Physically it simply means that
because of the presence of the spin gap it is now impossible for a
single particle to leave a tube since it would take away one spin
and thus destroy the spin gap. One would therefore need a critical
value of $t_\perp$ for this to happen. However, although the single
particle hopping is an irrelevant operator, it can generate relevant
perturbations at higher order\cite{giamarchi_book_1d}. Such relevant
perturbations corresponds to hopping between different tubes that do
not destroy the spin gap.

\begin{figure}
\begin{center}
\includegraphics[width=\largefig]{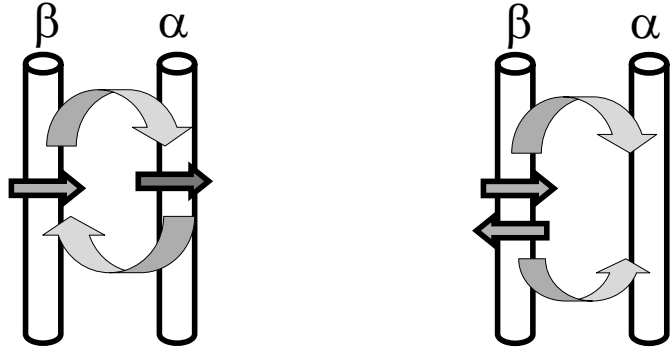}
\caption{Relevant coupling generated to second order by the single
particle intertube hopping. In order to avoid breaking the spin gap,
a correlated hop of two objects must be done simultaneously. Left:
particle-hole hopping. Since this term exchanges two particles on
two different tubes, it corresponds to a standard density-density or
spin-spin coupling term. This is the superexchange term. Right:
particle-particle hopping. This term allows a pair to jump from one
tube to the next and is thus analogous to the Josephson coupling
between superconductors. \label{fig:twotype}}
\end{center}
\end{figure}

Quite generally, as shown in \fref{fig:twotype}, two types of
relevant perturbations are possible when looking at the second order
term generated by (\ref{eq:hperp}).
The first one corresponds to particle-hole hopping between the
tubes, it is thus a density-density or spin-spin term, while the
second one is the hopping of two particles between between the tubes
and thus corresponds to a Josephson coupling. The simplest way to
analyze the terms generated by (\ref{eq:hperp}) is to use the
bosonization representation. The intertube tunnelling term becomes
\begin{eqnarray}
 H_\perp &\propto& -t_\perp
 \sum_{\sigma=\pm1,r=\pm1}\sum_{\langle\alpha,\beta\rangle} \int dx
 e^{\frac{i}{\sqrt2}[r\phi_{\rho,\alpha}(x)-\theta_{\rho,\alpha}(x) + s
 (r\phi_{\sigma,\alpha}(x)-\theta_{\sigma,\alpha}(x))]} \times \nonumber \\
 & & e^{\frac{-i}{\sqrt2}[r\phi_{\rho,\beta}(x)-\theta_{\rho,\beta}(x) + s (r\phi_{\sigma,\beta}(x)-\theta_{\sigma,\beta}(x))]}
\end{eqnarray}
Going to second order in (\ref{eq:hperp}), using the fact that
$\phi_\sigma$ is now ordered, and the fact that all operators
leading to exponential decay must be eliminated to get the leading
operator, one sees that a first surviving term is
\begin{equation} \label{eq:intsdw}
 H_{\SDW} = J_{\SDW} \sum_{\alpha,\beta} \int dx O_{\SDW_z,\alpha}^\dagger(x) O_{\SDW_z,\beta}(x)
\end{equation}
The value of the coupling constant can be obtained by standard
second order perturbation theory and is of the order of $J_{\SDW}
\sim t_\perp^2/\Delta_s$. This term is thus a standard superexchange
term coupling the antiferromagnetic spin modulation on two different
tubes. Quite naturally this term would tend to stabilize the
$\SDW_z$ instability that would develop in an isolated tube and lead
to a three dimensional ordered antiferromagnetic phase.

There is however another term that survives. This is a term where
two particles can hop from one tube to the other. It is of the form
\begin{equation} \label{eq:intts}
 H_{\TS} = J_{\TS} \sum_{\alpha,\beta} \int dx O_{\TS_z,\alpha}^\dagger(x) O_{\TS_z,\beta}(x)
\end{equation}
This term corresponds to the hopping of a pair of fermions, that are
in a triplet paring state from one tube to the other. Singlet
hopping is here cancelled because of the presence of the spin gap.
Because a superfluid pair has a global momentum of zero, the pair
hopping does not contain the oscillating $2\kf$ factor that is
present in the superexchange term (see (\ref{eq:chspinop})). Here
again the coupling constant is of the order of $J_{\TS} \sim
t_\perp^2/\Delta_s$.

\begin{figure}
\begin{center}
\includegraphics[width=\largefig]{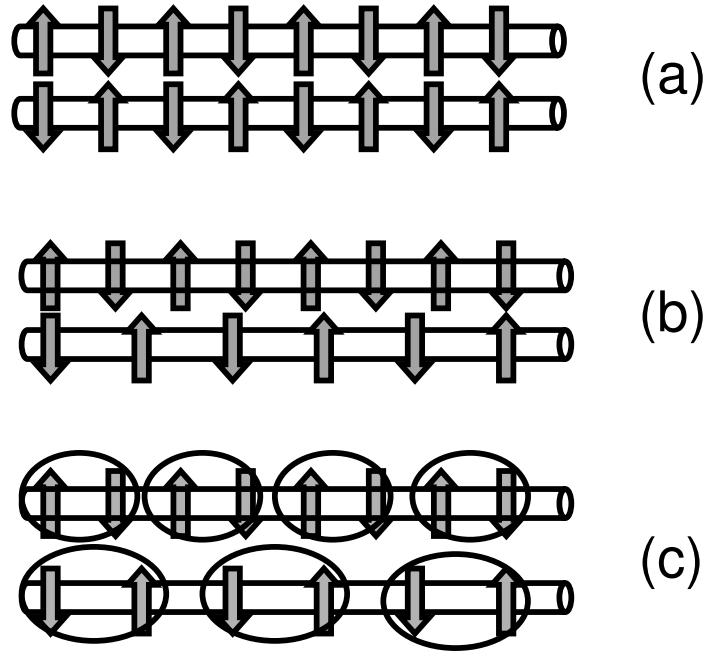}
\caption{Schematic diagrams for the true long range order induced by
inter-tube particle-pair or particle-hole hopping. a)
Antiferromagnetic order for $U>0$ and different velocities, with
equal Fermi momentum ({\it i.e.} equal number of fermions in each
tube). b) Frustration (incommensurability) to the antiferromagnetic
order between tubes when the Fermi momentum is different between the
tubes. c)  No such frustration effects when the order parameter
carries zero total momentum, namely for the superfluidity.
\label{fig:instab}}
\end{center}
\end{figure}

Both the superexchange and the Josephson term tend to stabilize
their corresponding type of order. What phase is realized normally
depends crucially on what is the leading one dimensional
instability. Usually the Fermi momenta of all tubes are the same and
there is no oscillatory factors in (\ref{eq:intsdw}). Since the
$\SDW$ order is dominant for the isolated tube, and both $J_\SDW$
and $J_\TS$ are of the same order of magnitude a simple RPA
treatment of the coupling term shows that the $\SDW$ phase is
stabilized. This is the situation depicted in \fref{fig:instab}(a).
This situation is normally the one realized in condensed matter
systems. It corresponds to the stabilization of a three dimensional
antiferromagnetic order for repulsive interactions. However the
situation can be much richer if the Fermi momenta of the different
tubes are different. Note that for cold atoms this is the rule
rather than the exception because of the presence of the parabolic
confining potential that makes each tube different depending from
its distance from the center. This situation can also be reinforced
artificially by adding an additional modulation of the optical
lattice. In that case one has $\kf^\alpha-\kf^\beta \ne 0$ which
means that the oscillatory factors remain in (\ref{eq:intsdw}). Such
factors considerably weaken the intertube superexchange. Physically
this means that the antiferromagnetic fluctuations on the
neighboring tubes are now incommensurate with each other and thus
cannot couple very well as depicted in \fref{fig:instab}(b). On the
other hand the Josephson term which is a $q\sim 0$ transfer between
the tube is not affected by such a difference of Fermi momentum and
remains unchanged, as schematically shown in \fref{fig:instab}(c).
One is now in a situation where this term can become
dominant\cite{cazalilla_fermions_1d} \emph{even} if the triplet
superfluid instability in the isolated tube is a subdominant
instability. One would thus be in a situation to stabilize a three
dimensional triplet superfluid phase.

This is a rather unique situation. From the theoretical point of
view such a stabilization of a subdominant instability by intertube
coupling is quite interesting and potentially relevant to other
situations as well. In particular, this mechanism is similar, albeit
for a triplet superconductor, to the one advocated to stabilize
singlet superconductivity for coupled
stripes\cite{arrigoni_ladder_superconductor}. Such cold atomics
system could thus be controlled systems in which to check for the
feasibility of such a mechanism. More directly it would be very
interesting to have a realization for a triplet superconductor.
Triplet superconductivity is indeed quite rare. Besides Helium 3,
strontium ruthenate is the only one candidate well identified in
condensed matter\cite{mackenzie2003}. Organic superconductors,
another system made of coupled chains, is also a potential candidate
for such unusual superconductivity\cite{ishiguro2002}. Cold atomic
gases of fermions could thus help shed a light on the mechanisms and
properties of unusual superconductivity in these very anisotropic
systems.

\section*{Acknowledgements}

M.A.C. is supported by \emph{Gipuzkoako Foru Aldundia} and MEC
(Spain) under grant FIS-2004-06490-C03-00, A.F.H. by EPSRC(UK) and
DIPC (Spain), and T.G. by the Swiss National Science Foundation
under MANEP and Division II.


\end{document}